\pdfoutput=1
\documentclass[amsmath,amssymb,aps,twocolumn,superscriptaddress]{revtex4-2}
\usepackage{amsmath}
\usepackage{amssymb}
\usepackage{bm}
\usepackage{graphicx}
\usepackage{epsf}
\usepackage{xcolor}
\usepackage{color}
\usepackage{graphicx}
\usepackage[version=4]{mhchem}
\usepackage[bookmarks=false]{hyperref}
\usepackage{siunitx}
\usepackage[english]{babel}
\usepackage[T1]{fontenc}
\usepackage{dcolumn}

\makeatletter
\def\maketitle{
\@author@finish
\title@column\titleblock@produce
\suppressfloats[t]}
\makeatother
\newcommand{\beginsupplement}{
    \setcounter{table}{0}
    \renewcommand{\thetable}{S\arabic{table}}
    \setcounter{figure}{0}
    \renewcommand{\thefigure}{S\arabic{figure}}
    \setcounter{equation}{0}
    \setcounter{section}{0}
    \renewcommand{\theequation}{S\arabic{equation}}
}

\begin{document}
\title{Bose-Einstein condensation of polaritons at room temperature in a GaAs/AlGaAs structure}

\author{Hassan Alnatah}
\affiliation{Department of Physics, University of Pittsburgh, 3941 O’Hara Street, Pittsburgh, Pennsylvania 15218, USA}

\author{Qi Yao}
\affiliation{Joint Quantum Institute, University of Maryland and National Institute of Standards and Technology, College Park,
Maryland 20742, USA}

\author{Qiaochu Wan}
\affiliation{Department of Physics, University of Pittsburgh, 3941 O’Hara Street, Pittsburgh, Pennsylvania 15218, USA}

\author{Jonathan Beaumariage}
\affiliation{Department of Physics, University of Pittsburgh, 3941 O’Hara Street, Pittsburgh, Pennsylvania 15218, USA}

\author{Ken West}
\affiliation{Department of Electrical Engineering, Princeton University, Princeton, New Jersey 08544, USA}

\author{Kirk Baldwin}
\affiliation{Department of Electrical Engineering, Princeton University, Princeton, New Jersey 08544, USA}

\author{Loren N. Pfeiffer}
\affiliation{Department of Electrical Engineering, Princeton University, Princeton, New Jersey 08544, USA}

\author{David W. Snoke}
\affiliation{Department of Physics, University of Pittsburgh, 3941 O’Hara Street, Pittsburgh, Pennsylvania 15218, USA}

\date{\today}

\begin{abstract}
We report the canonical properties of Bose-Einstein condensation of polaritons, seen previously in many low-temperature experiments, at room temperature in a GaAs/AlGaAs structure. These effects include a nonlinear energy shift of the polaritons, showing that they are not non-interacting photons, and dramatic line narrowing due to coherence, giving coherent emission with spectral width of 0.24 meV at room temperature with no external stabilization. This opens up the possibility of room temperature nonlinear optical devices based on polariton condensation.

\end{abstract}

\maketitle

\section{INTRODUCTION}

While many clear demonstrations of effects of Bose-Einstein condensation of polaritons have been seen in GaAs-based structures at low temperature  (e.g., Refs.~\cite{deng2002condensation,kasprzak2006bose,balili2007bose,abbarchi2013macroscopic,sanvitto2010persistent,lagoudakis2009observation}), it has generally been assumed that GaAs-based structures will not work for room-temperature condensates.  This intuition is largely based on the binding energy of the excitons in GaAs, which is roughly 10 meV in quantum wells \cite{bastard1982exciton}, well below the thermal energy scale at room temperature of $k_BT = 25.6$~meV. However, it is not the case that polaritons can only be formed when excitons are fully stable bound states. Rather, the exciton-polariton is a quasiparticle state of the system that can coexist with free carriers and uncoupled excitons.

When there is light-matter coupling between excitons and photons, all that is needed for the existence of polaritons is an exciton resonance that lies reasonably near to the energy of the photon mode. At room temperature, the exciton resonance in GaAs is broadened substantially due to phonon scattering, which tends to reduce the coupling between excitons and photons, but it does not completely destroy the coupling. A simple three-level model for our GaAs-based microcavity structures which includes heavy-hole excitons, light-hole excitons and photons can be written as
\begin{equation}
H = 
\begin{pmatrix}
 E_p + i\Gamma_p& \Omega/2 & \Omega/2\\ 
\Omega/2 & E_\mathrm{hh} + i\Gamma_{\mathrm{hh}} & 0\\ 
\Omega/2 & 0 & E_\mathrm{lh} + i\Gamma_{\mathrm{lh}}
\end{pmatrix},
\end{equation}
where $\Gamma_p$ gives half width at half maximum of the photon linewidth broadening, here about $\hbar/(100~{\rm ps}) \sim .01~{\rm meV}$,  and  
$\Gamma_{\mathrm{hh}}$ and $\Gamma_{\mathrm{lh}}$ give the heavy-hole and light-hole exciton broadening, here set to 7.3~meV and 24.6~meV, respectively, based on measurements of the bare quantum well photoluminescence (PL) at room temperature (given in the Supplemental information). For a moderate exciton-photon coupling of $\Omega = 4.5$~meV, and our experimentally measured cavity photon and exciton energies, this gives a well-defined lower polariton line, with a half width of 0.69~meV and an exciton fraction of 8.5\%

In the experiments reported here, we used a GaAs/AlGaAs microcavity structure very similar to those of previous experiments \cite{alnatah2024coherence,alnatah2024critical,sun2017bose}, but in which the cavity photon energy falls very near the quantum well exciton energy at room temperature. We see all of the expected behavior for a polariton condensate, as seen previously in low-temperature experiments. This system has some aspects in common with pure photon condensates \cite{klaers2010bose,pieczarka2023boseeinstein}, but also important differences that arise from the interactions of the particles---in particular, strong nonlinearity. This opens up the possibility of optical transistor devices that operate at room temperature. The coherent emission also has an ultra-narrow linewidth, without any active stabilization of the mode. 
\section{EXPERIMENTAL OBSERVATIONS}
The microcavity in this study consisted of a total of 12 GaAs quantum wells with AlAs barriers embedded within a distributed Bragg reflector (DBR). The DBRs are made of alternating layers of AlAs and Al$_{0.2}$Ga$_{0.8}$As. The quantum wells are in groups of 4, with each group placed at one of the three antinodes of the $3\lambda/2$ cavity. The large number of DBR periods gives the cavity a high Q-factor, resulting in a cavity lifetime of $\sim 135$ ps and a polariton lifetime of $\sim 270$ ps at resonance \cite{steger2015slow}. 
\par
\begin{figure*}
\centering
\includegraphics[width=0.7\textwidth]{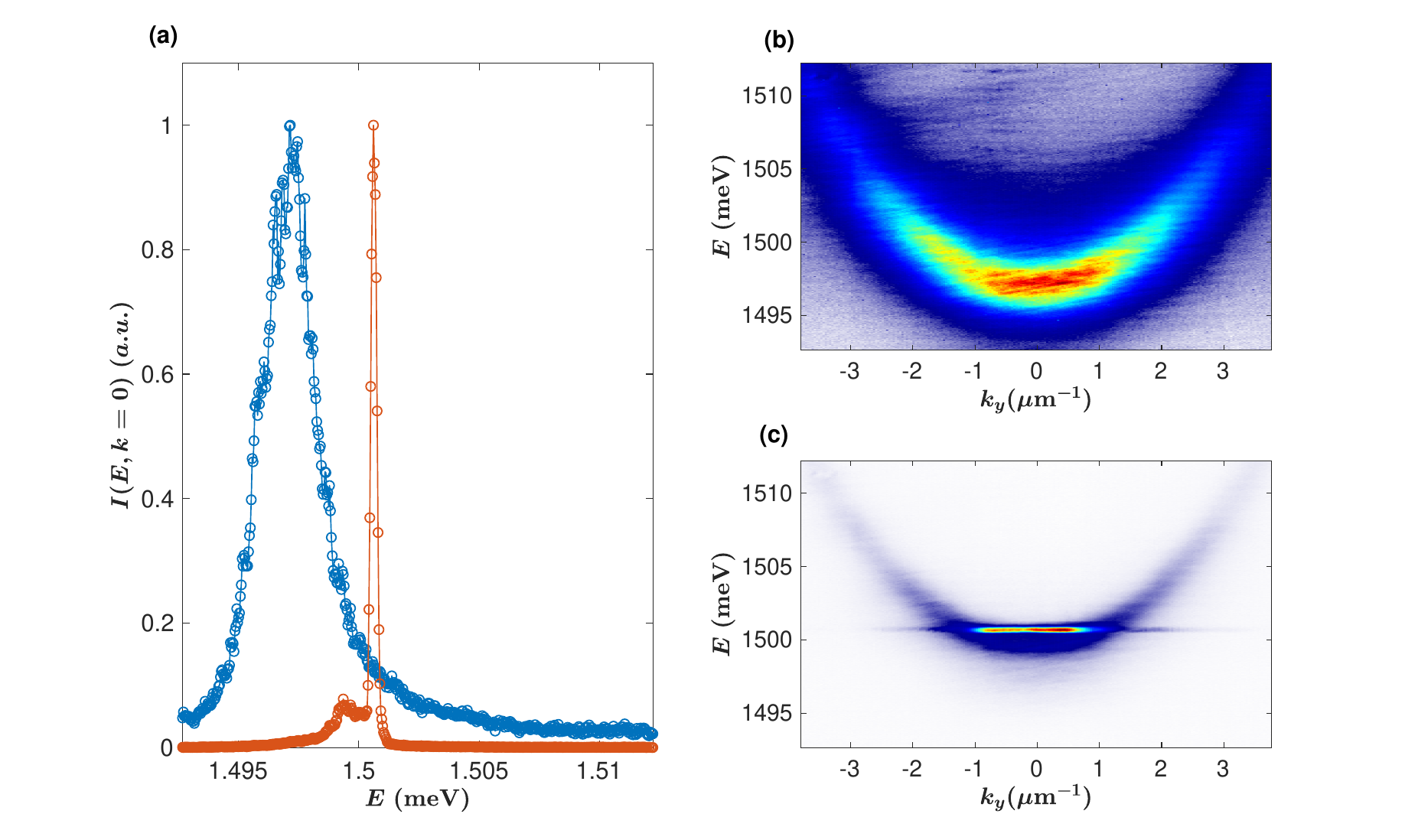}
\caption{\textbf{Ultralow linewidth.}(a) The intensity at $k=0$ at $P/P_{th} = 0.8$ (blue) and $P/P_{th} = 1.34$. The linewidth is extracted by fitting a Lorenzian, giving a linewidth of $ 2.5 \;\mathrm{meV}$ for the blue curve and $ 0.24 \;\mathrm{meV}$ for the red curve. (b) the polariton energy dispersion corresponding to blue curve in (a). (b) the polariton energy dispersion corresponding to red curve in (a). }
\label{narrow_linewidth}
\end{figure*}
With the sample at room temperature, we generated polaritons by pumping the sample non-resonantly with an M Squared wavelength-tunable laser, which was tuned to a reflectivity minimum, about 174 meV above the lower polariton resonance (740 nm). The laser focus was a Gaussian with full width at half maximum (FWHM) $\sim 45~ \mathrm{\mu m}$; to reduce heating of the sample, we modulated the pump laser using an optical chopper with a duty cycle of 1.7$\%$ and pulses of duration 41.6 $\mathrm{\mu s}$, which is very long compared to the dynamics of the system. The non-resonant pump created electrons and holes, which scattered down in energy to become polaritons.  The photoluminescence (PL) was collected using a microscope objective with a numerical aperture of 0.75 and was imaged onto the entrance slit of a spectrometer. The image was then sent through the spectrometer to a CCD camera for time-integrated imaging. 
\begin{figure}
\centering
\includegraphics[width=0.7\columnwidth]{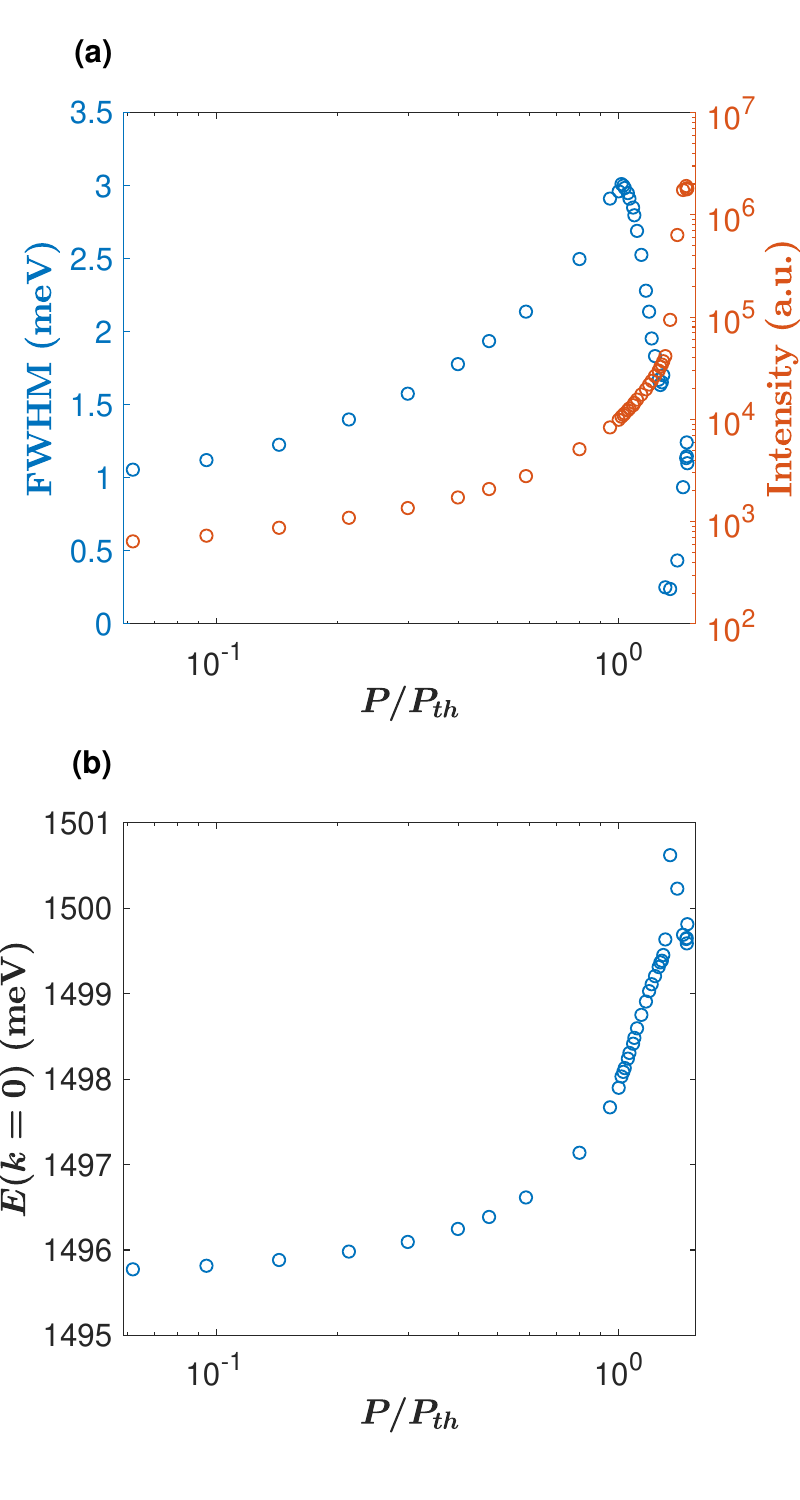}
\caption{\textbf{Blue shift and linewidth narrowing.} (a) Full width at half max at $k=0$ and the intensity as a function of the pump power. (b) the blue shift at $k=0$ as a function of the pump power.}
\label{linewidth}
\end{figure}
\par
Polariton condensation was observed as we increased the pump power across the BEC phase transition. The pump power needed was only about a factor of two higher than in the low temperature experiments in this type of structure, corresponding to a peak optical intensity of $1.5\times 10^4$ W/cm$^2$, corresponding to an injected carrier density of approximately $5.7\times 10^{22} \; \mathrm{cm^{-2}\; s^{-1}}$. As seen in Fig~\ref{narrow_linewidth}(a), the linewidth becomes ultra-narrow when the pump power is above the condensation threshold, reaching a linewidth of $ 0.24 \;\mathrm{meV}$ (Cf.~Ref.~\cite{wangultra}, which reported a linewidth of 0.7 meV.)
Fig.~\ref{narrow_linewidth} (b-c) show that the spectral line narrowing is accompanied by momentum-space narrowing as expected for a Bose condensate.

A sharp nonlinear increase of the intensity by two orders of magnitude is also observed near the threshold of condensation. This nonlinear increase in intensity is accompanied with a factor of $\sim ~10$ decrease in linewidth and significant energy blue shift (Fig.~\ref{linewidth}), all of which are hallmarks of polariton Bose–Einstein condensation \cite{kasprzak2006bose,balili2007bose}. The blue shift in particular shows that there is a nonlinear interaction of the polaritons with excitons, which are produced by the non-resonant pump. This confirms that the polaritons have a significant exciton component. At the highest density, there is a slight red shift, which can be due to either lattice heating or depletion of the exciton cloud in the region of the condensate \cite{estrecho2018single}.

\begin{figure}
\includegraphics[width=0.5\columnwidth]{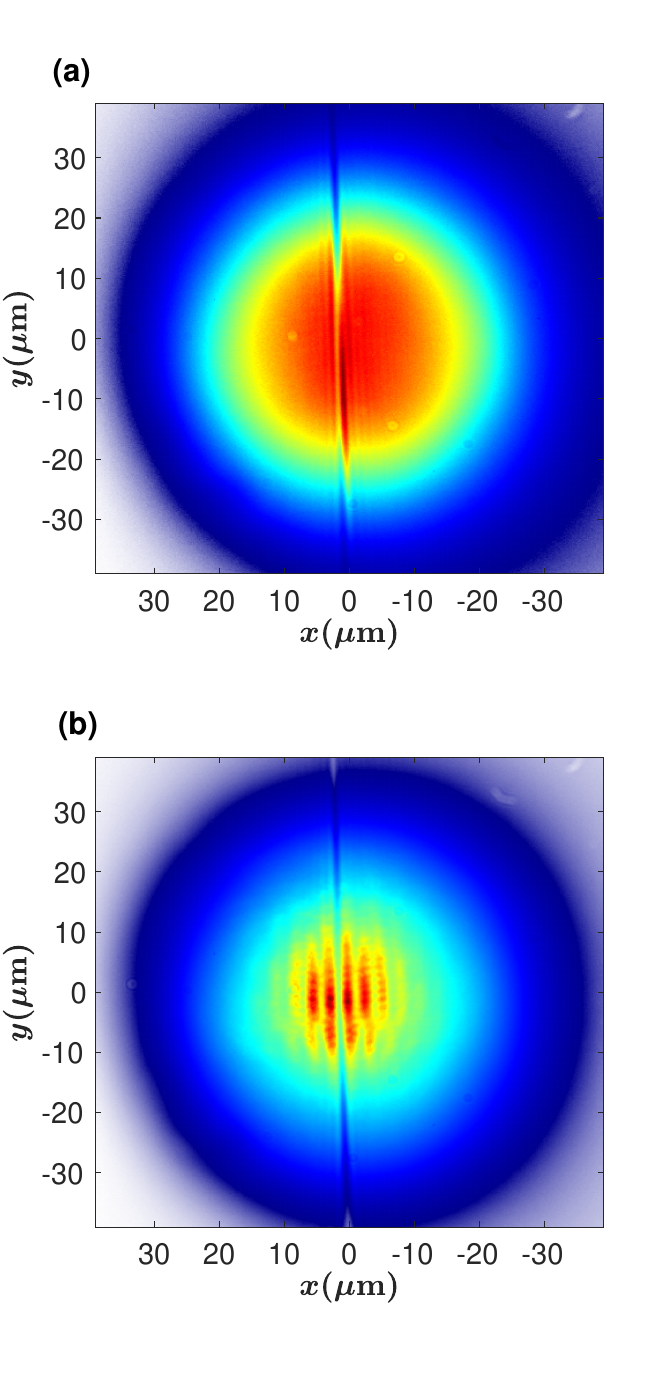}
\centering
\caption{\textbf{Coherence of polaritons.} Real-space interference of the polaritons, recorded by superposing the spatial image of the emission with its mirror-image. (a) $P/P_{th} = 0.17$ and (b) $P/P_{th} = 1.22$}
\label{rs_interference}
\end{figure}

In addition, we measured the coherence of the polariton by interfering the electric field emitted from the cavity $E(x,y,t_0)$ with its mirror image $E(-x,y,t_0)$ using Michelson interferometry. We observed extended spatial coherence and an increases of the visibility as the pump power is increased (Fig. \ref{rs_interference}).
\par
We found that for photonic detunings, the polaritons tend to self-trap into the center of the pump spot for densities above the threshold of condensation, an effect which has been well observed at low temperature \cite{estrecho2018single}.
This leads to a multimode behavior of the condensate when it has mostly photonic character. To minimize this self-trapping effect, the data reported here were taken at higher exciton fraction. The exciton fraction can be chosen by moving to different locations on the sample with different cavity photon energy.
\begin{figure*}
\includegraphics[width=0.7\textwidth]{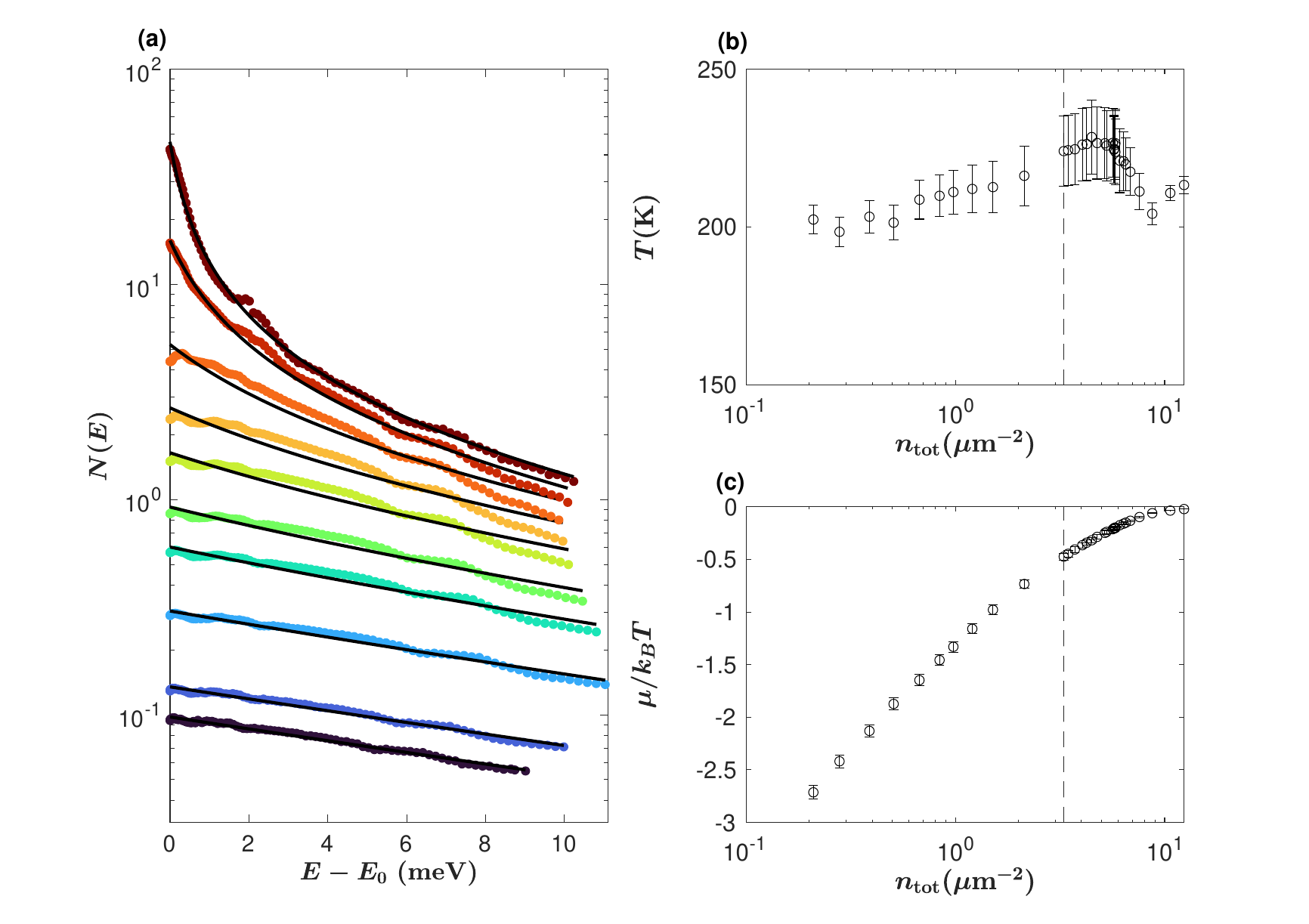}
\caption{\textbf{Thermalization of polaritons.} (a) Occupation number of the polaritons as a function of energy. The solid lines are best fits to the equilibrium Bose-Einstein distribution. (b) the effective temperature and (b) the reduced chemical potential of the polaritons obtained from the fits to the Bose-Einstein distribution. The vertical dashed line in (b-c) denotes the critical density, defined as the total density of the polaritons at $P/P_{th} = 1$. }
\label{NvsE}
\end{figure*}
\par
To study thermalization of the these polaritons, we used angle-resolved imaging to obtain the spectral function $I(k,E)$, as done in previous works \cite{sun2017bose,alnatah2024coherence} and described in the Supplementary Material. This spectral function was then converted to an occupation number $N(E)$ using a single efficiency factor. Figure \ref{NvsE} shows the distribution function of polaritons for different pump powers.  We used a circular real space filter with a radius of $\sim 22\;\mathrm{\mu m}$ to collect light from a fairly homogeneous region of the PL. The measured occupation number was then fit to the Bose-Einstein distribution, given by
\begin{equation}
\begin{aligned}
N\left ( E \right ) = \frac{1}{e^{\left (E-E(0)-\mu  \right )/k_{b}T}-1},
\label{eq:BE}
\end{aligned}
\end{equation}
where $T$ and $\mu$ are fit parameters, which give the temperature and chemical potential of the polaritons, $E$ is the polariton energy, and $E(0)$ is the polariton ground state energy at $k  = 0$, which shifts to higher energy as the density increases, due to many-body renormalization \cite{sun2017bose}. We find that polaritons are well described by the Bose-Einstein distribution as shown by the fits in Fig.~\ref{NvsE}(a). At low density, the Bose-Einstein distribution becomes a Maxwell-Boltzmann distribution, which is a straight line in semi-log scale. At high density, the Bose-Einstein statistics become important, giving rise to an upturn at the ground state energy $E(k=0)$. 
\par
Interestingly, the polaritons equilibrate to a temperature $T\sim 220\; \mathrm{K}$ lower than the reservoir $T\sim 293\; \mathrm{K}$. This was also seen in Ref. \cite{pieczarka2023boseeinstein} for non-interacting photons. Although it is surprising, we know that the polaritons are generally decoupled from the phonon and bath temperature; at low temperature, the polaritons are typically 20 K {\em above} the bath temperature. Cooling will occur whenever the most frequent down-conversion of excitons into polaritons occurs into low-energy polariton states.
A recent theoretical work \cite{shishkov2022exact} predicted  a simulated cooling effect, but as seen in Figure 4(b), we observe effective polariton temperatures below room temperature even at very low pump density.
\par
The critical density of condensation at room temperature can be estimated by relating the average distance between the particles to the de Broglie wavelength $a \sim \lambda_{th}$, which gives
\begin{equation}
\begin{split}
n_c &\sim \left [ \frac{h}{\sqrt{2\pi m k_{B}T}} \right ]^{-2}\approx 1.6\;\si{\mu m^{-2}},
\end{split}
\end{equation}
where $m = (3.96\pm 0.95)\times 10^{-5}m_e$ is the mass of the polaritons (measured from data like that of Fig.~\ref{narrow_linewidth}(b)), and $T\sim 220 \; \mathrm{K}$ is the temperature. This is within the uncertainty of the directly measured density using a calibrated light source to establish the absolute photon emission rate, which gives $n = 2.97\pm 1.75\;\si{\mu m^{-2}}$ for the polaritons at $P/P_{th} = 1$.

\section{Conclusions}

Many material systems such as transition metal dichalcogenides \cite{liu2019room,zhao2021ultralow}, perovskites \cite{su2020observation,su2018room}, and organics \cite{plumhof2014room,dusel2020room,wei2022optically} have been proposed and studied for room temperature Bose condensation for the purpose of nonlinear coherent device applications. Our observation of room temperature BEC here result shows that the same effects can occur in the well-studied system of GaAs and related III-V semiconductors, which can have extremely high quality and uniformity. 

There are many commercial vertical-cavity surface-emitting lasers (VCSELs) with similar design to our structure, including the recent devices used for photon condensation \cite{pieczarka2023boseeinstein}.  We believe that similar designs should be able to show the nonlinear behavior reported here, allowing for transistor-like operation \cite{zasedatelev2019all} 

\section{Acknowledgements}
The experimental work at Pittsburgh and sample fabrication at Princeton were supported by the National Science Foundation, grant  DMR-2306977. 
\clearpage
\bibliography{references.bib}
\clearpage
\title{Supplementary Information for: Bose-Einstein condensation of polaritons at room temperature in a GaAs/AlGaAs structure}
\date{\today}
\maketitle
\beginsupplement
\section{Density Calibration}
Here, we compare two independent methods to calibrate the density of the polaritons. The first method uses a photon calibration for our setup that relates the CCD count on the camera to the number of photons detected. This calibration was created by tuning the wavelength of the M Squared laser close to the polariton emission wavelength (810 nm). We then placed a mirror at the sample plane, which reflected the laser to the CCD camera through the same optical set up that was used in the experiment. The total CCD count is related to the number of photons through the equation
\begin{equation}
I_{\mathrm{CCD}} =\frac{1}{\eta}\frac{N_{\mathrm{ph}}}{\Delta t},
\end{equation}
where $\eta$ is the efficiency factor, $\Delta t$ is the integration time of the camera and $N_{\mathrm{ph}}$ is the number of photons detected by the camera. This can then be related to the number of photons by measuring the power of the laser at the sample plane. The total number of photons the camera detects during an integration time  $\Delta t$ is given by 
\begin{equation}
N_{\mathrm{ph}} = \frac{P\Delta t}{hc/ \lambda},
\end{equation}
where $P$ is the measured power of the laser at the sample plane, $h$ is Planck constant, $c$ is the speed of light and $\lambda$ is the wavelength of the laser. The efficiency factor is therefore given by
\begin{equation}
\eta = \frac{P\lambda}{hc} \frac{1}{I_{\mathrm{CCD}}}
\end{equation}
This efficiency factor allows us to relate the counts on the camera to the number of photons sent to the camera during a time $\Delta t$. The total density of the polaritons for a given $I_{\mathrm{CCD}}$ count is then given by:
\begin{equation}
n_{\mathrm{tot}} = \frac{\eta I_{\mathrm{CCD}}\tau}{A_{\mathrm{obs}}},
\end{equation}
where $A_{\mathrm{obs}}$ is observation area on the sample, $\tau$ is the lifetime of the polaritons.  To account for the laser chopping for the PL, we divide by the duty cycle $d = 1.7\%$, which gives
\begin{equation}
n_{\mathrm{tot}} = \frac{\eta I_{\mathrm{CCD}}\tau}{A_{\mathrm{obs}}d},
\end{equation}
\par
The second method that we used to calibrated the density of the polaritons is by fitting the extracted occupation number of the polaritons to the Bose–Einstein distribution. This was done by fitting multiple occupation numbers at different pump powers to the Bose–Einstein distribution using $T$ and $\mu$ as fit parameters and a single efficiency factor $\eta$. This then constraints the efficiency factor by minimizing the mean-squared error in the fit process. Therfore, for a set of $n$ number of occupation numbers each corresponding to a different pump power, we have a total of $2n+1$ free parameter, i.e. $n$ temperature parameters, $n$ chemical potential parameters and one single efficiency factor. 
\par
The two methods give consistent results. At the threshold of condensation, the photon calibration methods gives a density of $n = 2.97\pm 1.75\;\si{\mu m^{-2}}$ while the best fit to the Bose-Einstein distribution gives a density of $n = 3.3\pm 0.49\;\si{\mu m^{-2}}$.

\section{Extracting the occupation}
\begin{figure}
\includegraphics[width=0.9\columnwidth]{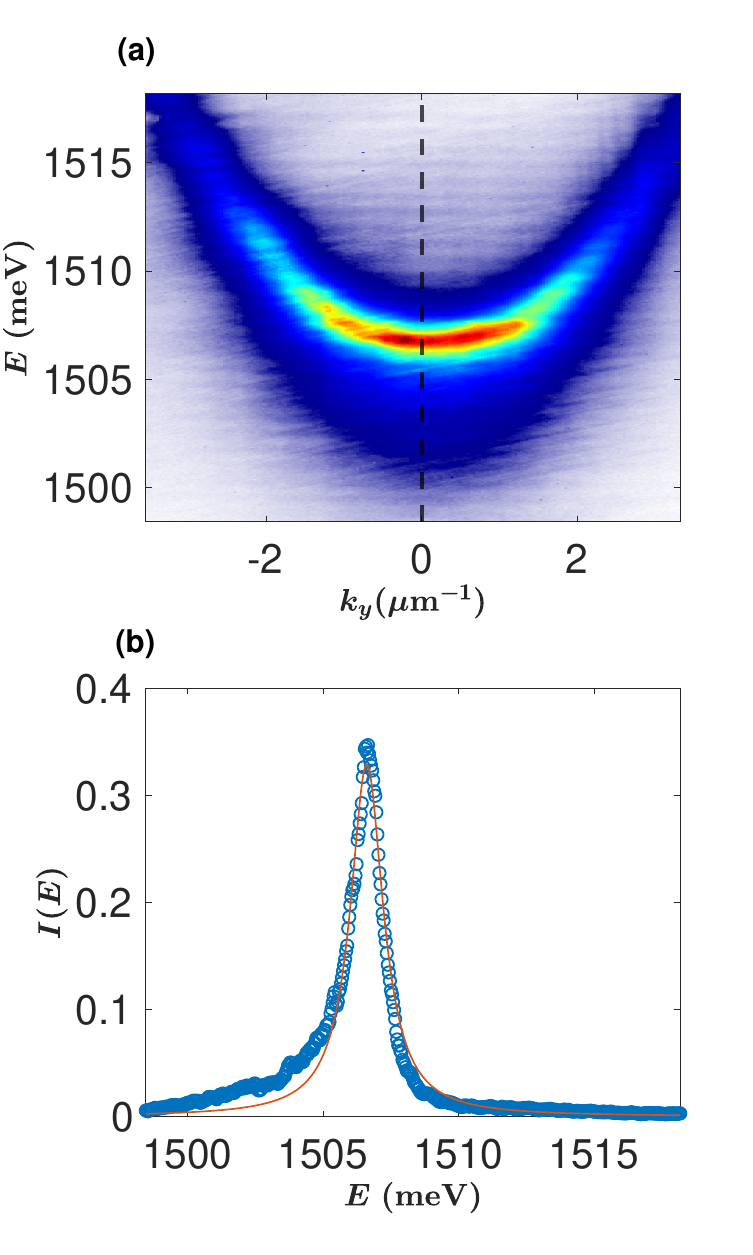}
\centering
\caption{\label{E_k_image} \textbf{Method of extracting the occupation.} (a) A typical energy dispersion of the thermalized polaritons. (b) A vertical slice in (a) at $k_{y} = 0$ showing the CCD counts. The red line is a fit to a Lorentzian function predicting an energy $1.5066$ meV for the polaritons at $k_{y} = 0$. The occupation number for this energy is proportional to the integral of the $I(E)$ curve.}
\end{figure}

As described in the main text, we used angle-resolved imaging to measure the spectral function $I(k,E)$ of the lower polariton. Figure \ref{E_k_image} shows a representative $I(k,E)$ of the lower polaritons at low density. The occupation number is extracted by taking vertical slices at each $k$ value to obtain $I(E)$ for each $k$ slice.This $I(E)$ curve is then fit with a Lorentzian function to extract the polariton energy for each $k$ slice (see for example Fig.~\ref{E_k_image}(b)). The occupation $N$ for each $k$ slice is then related to $I(E)$:
\begin{equation}
\begin{aligned}
N(k_{i}) = \eta \tau(k_{i})\int \mathrm{d}E\; I(k_{i},E),
\end{aligned}
\end{equation}
where $\tau(k)$ is the $k$-dependent radiative lifetime, and $\eta$ is an overall constant that can be determined by minimizing the mean-squared error in fitting a set of distributions $N(E)$. Since the polaritons are photonic and in the weak coupling regime, the lifetime is not a strong function of $k$ and can be taken as a constant $\tau(k)\sim \tau$. 

\section{Exciton energy and linewidth}
\begin{figure}
\includegraphics[width=0.9\columnwidth]{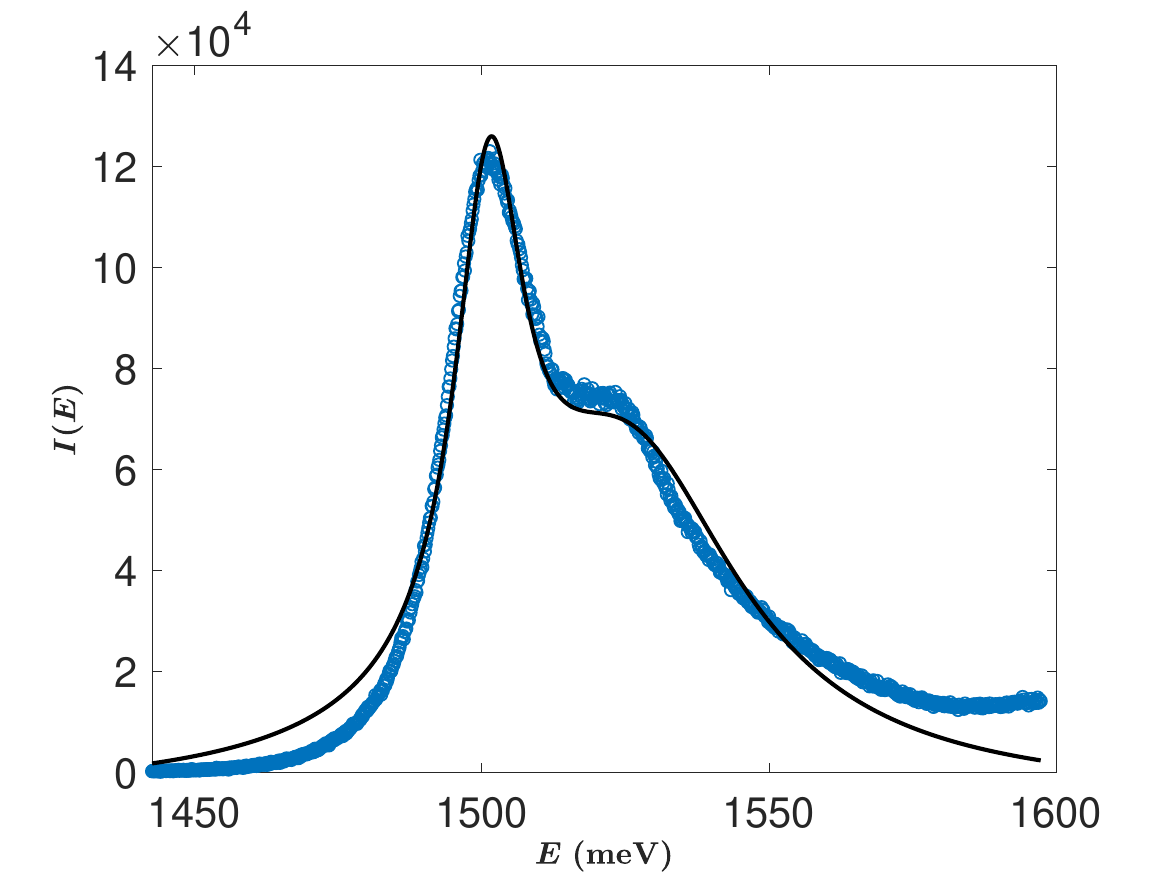}
\centering
\caption{\textbf{Exciton energy.}  The intensity $I(k=0,E)$ of GaAs excitons at room temperature. The solid black is a two-Lorentzian fit.}
\label{exciton_energy}
\end{figure}

We have measured the exciton PL at room temperature for a sample with bare GaAs quantum wells of the same width as in our microcavity sample, grown on a substrate at Princeton using the same method. Figure \ref{exciton_energy} shows the PL $I(k,=0,E)$ using a non-resonant excitation.  We see two-peaks-- the low energy peak corresponds to the heavy-hole exciton and the high energy peak is the light-hole exciton. We performed a two-Lorentzian fit to extract the energy and linewidth. The heavy-hole exciton has an energy of $E_0 = 1501.3\;\mathrm{meV}$ and a linewidth of $14.6~\mathrm{meV}$, while the light-hole exciton has an energy of $E_0 = 1525.2\;\mathrm{meV}$ and a linewidth of $49.1~\mathrm{meV}$. Interestingly, the ratio of the integral of the two Lorentzians is  $\sim 0.43$, which is approximately equal to ratio of heavy-hole excitons to light-hole exciton given by the thermal Boltzmann factor $\exp (-\Delta E/k_{B}T)\sim 0.4$, where $\Delta E \approx 23.9 \;\mathrm{meV}$ and $T\approx 300\;\mathrm{K}$.

\section{Polariton linewidth and energy blue-shift}

\begin{figure*}
\includegraphics[width=0.7\textwidth]{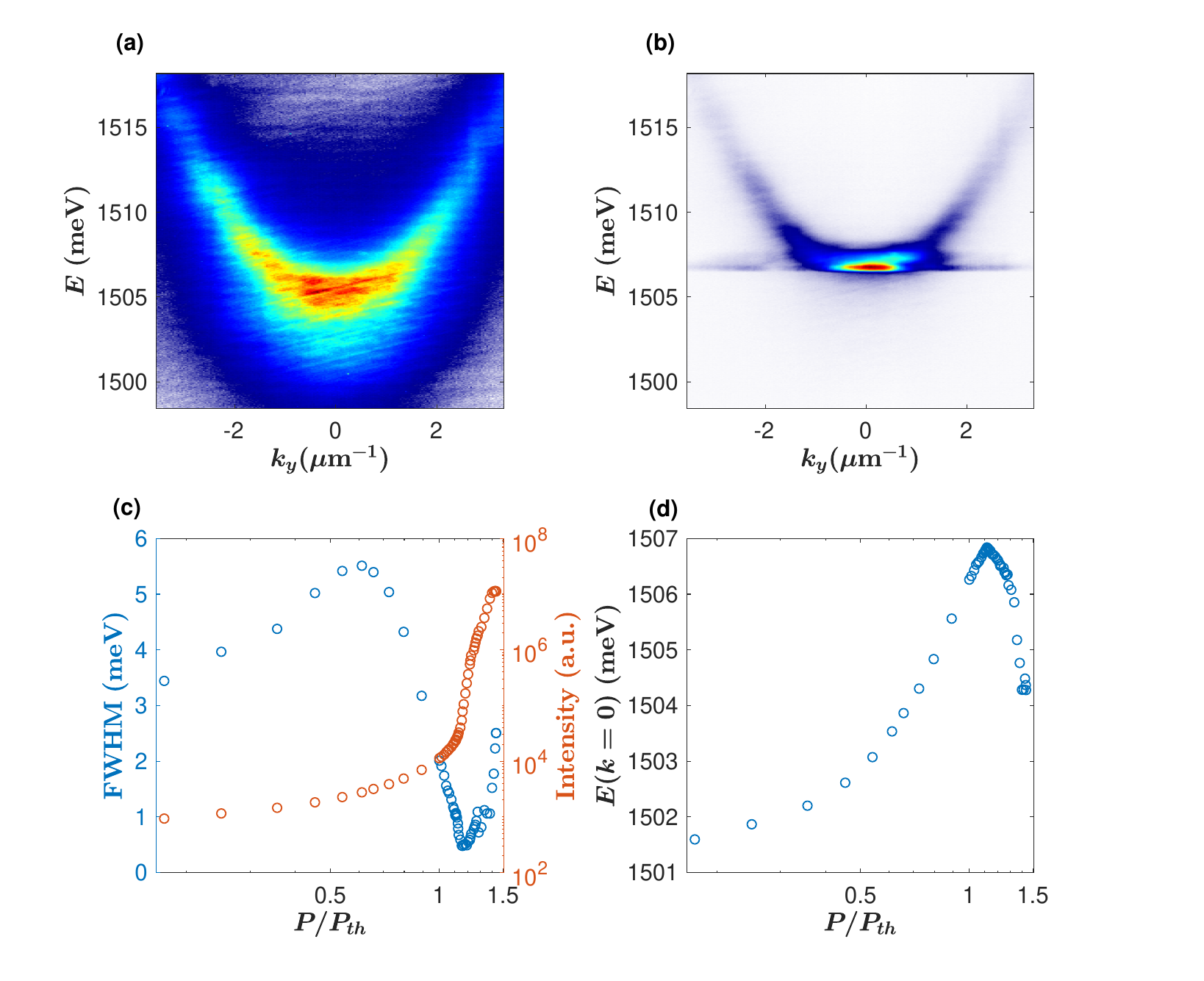}
\centering
\caption{\textbf{Blue shift and linewidth narrowing.} (a) the polariton energy dispersion corresponding at $P/P_{th} = 0.73$ and (b) at $P/P_{th} = 1.16$ .(c) Full width at half max at $k=0$ and the intensity as a function of the pump power. (d) the blue shift at $k=0$ as a function of the pump power.}
\label{linewidth_NE}
\end{figure*}

Figure 4 in the main text is from a different location on the sample than the data of Figs~1-3, at which the polaritons have slightly more photonic character. Figure \ref{linewidth_NE} shows the data of the polaritons corresponding to Fig. 4 in the main text. The behavior is quite similar to that of Figs.~1-3 of the main text; in general, we find that the effects reported are easily reproducible over a wide area of the sample. The minimum linewidth we observed for this location on the sample is $0.48\;\mathrm{meV}$. The polaritons remain mostly thermal but we have observed in some cases a secondary condensate mode can appear. This secondary mode is typically much weaker in intensity than the main mode at $k=0$. Therefore, the polaritons remain in quasi-equilibrium. 
\section{Three level model}
In this section, we use the three level model discussed in the main paper to simulate $I(E,k)$ for the experimental parameters. A simple three-level model for our GaAs-based microcavity structures which includes heavy-hole excitons, light-hole excitons (see Fig.~\ref{exciton_energy}) and photons can be written as
\begin{equation}
H(k) = 
\begin{pmatrix}
 E_p + i\Gamma_p& \Omega/2 & \Omega/2\\ 
\Omega/2 & E_\mathrm{hh} + i\Gamma_{\mathrm{hh}} & 0\\ 
\Omega/2 & 0 & E_\mathrm{lh} + i\Gamma_{\mathrm{lh}}
\end{pmatrix},
\end{equation}
where $\Gamma_p$ gives half width at half maximum of the photon linewidth broadening, here about $\hbar/(100~{\rm ps}) \sim .01~{\rm meV}$, and $\Gamma_{\mathrm{hh}}$ and $\Gamma_{\mathrm{lh}}$ give the heavy-hole and light-hole exciton broadening, here set to 7.3~meV and 24.6~meV, respectively. The exciton energies are $E_\mathrm{hh} = 1501.3\; \mathrm{meV}$ and $E_\mathrm{hh} = 1525.2\; \mathrm{meV}$. The energies and linewidths of the light and heavy excitons are extracted from Fig.~\ref{exciton_energy}. The cavity photon energy is given by
\begin{equation}
E_p =  E_{p0} + \frac{\hbar^2 k^2}{2m_{p}},   
\end{equation}
where cavity zero energy is taken to be $E_{p0} = 1.4981 \;\mathrm{meV}$ and the cavity mass is set to $m_p = 4.7\times 10^{-5}\; m_e$, where $m_e$ is the mass of the electron in vacuum. 
\par
We diaonalized the Hamiltonian for each $k$ value to find the real and imaginary energies and the linewidths of the lower and middle polaritons. This then allowed us to compute a simulated $I(k,E)$ image by adding two Lorentzians together with the fit center energies and widths from the eigenenergies at each $k$. Therefore, the intensity of the lower polariton is given by
\begin{equation}
I_{LP}(E) =\frac{A}{\pi}\frac{\frac{1}{2}\Gamma_{LP}}{(E-E_{LP})^2+\left (  \frac{1}{2}\Gamma_{LP}\right )^2}, 
\end{equation}
where $\Gamma_{LP}$ is the full width at half max of the lower polariton and $E_{LP}$ is the lower polariton energy. To insure that the polaritons are in an equilibrium distribution, we enforce that 
\begin{equation}
A\int\mathrm{d}E\;I_{LP}(E) = e^{-(E_{LP}-E_{0})/k_B T},
\end{equation}
which implies that 
\begin{equation}
A = \frac{e^{-(E_{LP}-E_{0})/k_B T}}{\int\mathrm{d}E\;I_{LP}(E)}.
\end{equation}
To account for the radiative lifetime of the polaritons, we multiply by the photon fraction, which then gives $N_{LP}(E) = AC_{LP}(k)I_{LP}(E)$. Similarly, for the middle polariton, we have
\begin{equation}
I_{MP}(E) =\frac{A}{\pi}\frac{\frac{1}{2}\Gamma_{MP}}{(E-E_{MP})^2+\left (  \frac{1}{2}\Gamma_{MP}\right )^2}.  
\end{equation}
Accounting for the thermalization factor and the radiative lifetime, $N_{MP}(E) = BC_{MP}(k)I_{LP}(E)$, where $B$ is given by
\begin{equation}
B = \frac{e^{-(E_{MP}-E_{0})/k_B T}}{\int\mathrm{d}E\;I_{MP}(E)}.
\end{equation}
The total intensity is then given by
\begin{equation}
I_{\mathrm{tot}} = N_{LP}(E) + N_{MP}(E).
\end{equation}

\begin{figure}
\includegraphics[width=0.9\columnwidth]{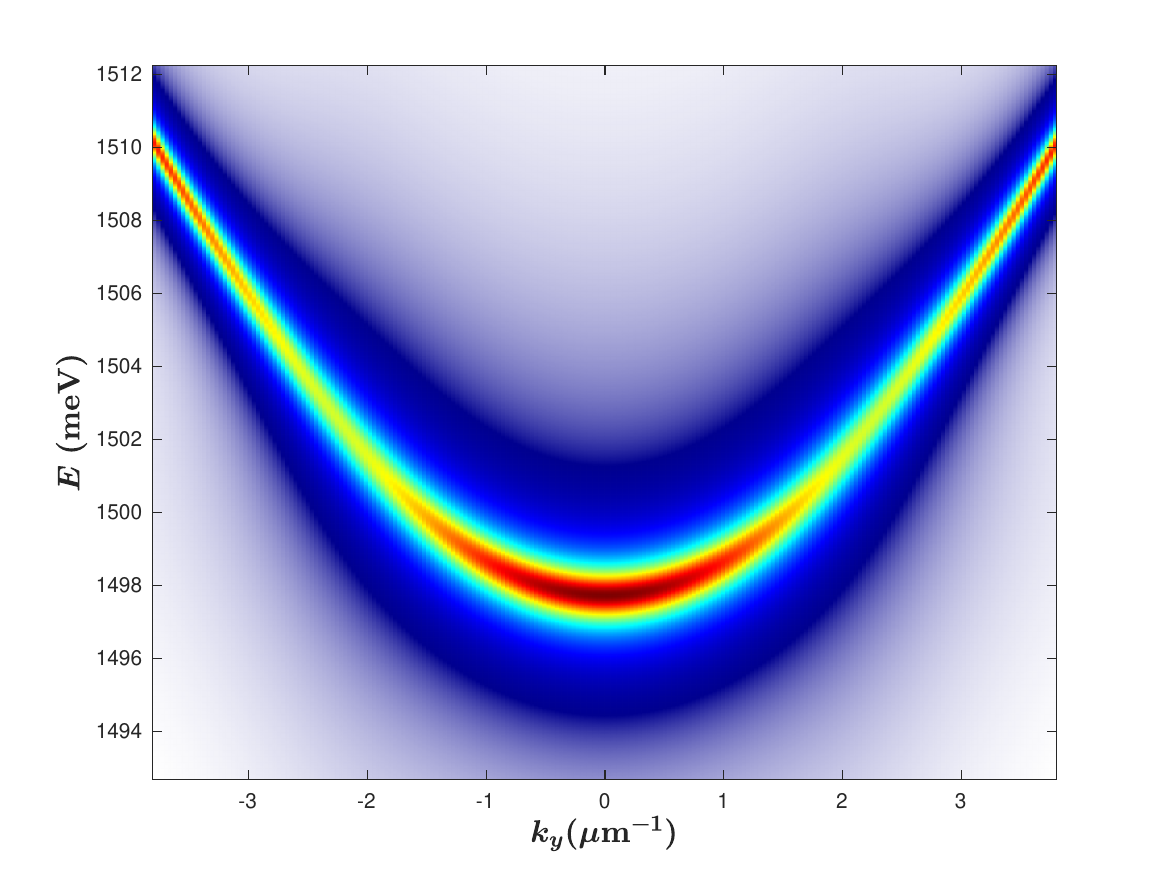}
\centering
\caption{\textbf{Simulated energy dispersion.}  The energy dispersion from the three level model using $\Omega = 4.5 \;\mathrm{meV}$}
\label{simulated_im}
\end{figure}
Figure \ref{simulated_im} shows the simulated $N(k,E)$ image. For this fit, the total exciton fraction (light hole plus heavy hole) is $8.5 \%$ at $k=0$.
\end{document}